# Additive Manufacturing of Ceramics from Preceramic Polymers: A Versatile Stereolithographic Approach Assisted by Thiol-Ene Click Chemistry

Xifan Wang[1]*, Franziska Schmidt[1], Dorian Hanaor[1], Paul H. Kamm[2], Shuang Li[3], and Aleksander Gurlo[1]

*correspondence to: xifan.wang@ceramics.tu-berlin.de

[1] Fachgebiet Keramische Werkstoffe / Chair of Advanced Ceramic Materials, Technische Universität Berlin, Hardenbergstr. 40 BA3, Berlin, 10623, Germany.

[2] Helmholtz-Zentrum Berlin für Materialien und Energie, Institut für Angewandte Materialforschung / Institute of Applied Materials, Hahn-Meitner-Platz 1, Berlin, 14109, Germany

[3] Functional Materials, Department of Chemistry, Technische Universität Berlin, Hardenbergstr. 40 BA2, Berlin, 10623, Germany

**Abstract**

Here we introduce a versatile stereolithographic route to produce three different kinds of Si-containing thermosets that yield high performance ceramics upon thermal treatment. Our approach is based on a fast and inexpensive thiol-ene free radical addition that can be applied for different classes of preceramic polymers with carbon-carbon double bonds. Due to the rapidity and efficiency of the thiol-ene click reactions, this additive manufacturing process can be effectively carried out using conventional light sources on benchtop printers. Through light initiated cross-linking, the liquid preceramic polymers transform into stable infusible thermosets that preserve their shape during the polymer-to-ceramic transformation. Through pyrolysis the thermosets transform into glassy ceramics with uniform shrinkage and high density. The obtained ceramic structures are nearly fully dense, have smooth surfaces, and are free from macroscopic voids and defects. A fabricated SiOC honeycomb was shown to exhibit a significantly higher compressive strength to weight ratio in comparison to other porous ceramics.

## 1. Introduction

In industrial applications, ceramics are most commonly formed by slip casting, injection molding and various pressure-assisted techniques. As all of these methods rely on the usage of molds, the shape of ceramic parts is restricted to relatively simple geometries. The machining of complex shapes is further hindered by the high hardness and brittleness of ceramic materials. Additive manufacturing (AM) heralds a new era in the fabrication of ceramics. As AM technologies are based on the layer by layer consolidation of ceramic powders, they neither require molds nor are they limited by the hardness and brittleness of ceramics. This in turn allows for the fabrication of complex parts that utilize the exceptional properties of ceramics such as high strength, thermal stability and chemical resistance [1,2]. Traditional AM approaches towards ceramics rely on powder- and slurry-based technologies including stereolithography of ceramic slurries containing UV-curable photopolymers [3,4], laser sintering of ceramic powder beds [5,6] and binder jetting, where a liquid organic binder is injected into a ceramic powder bed [7,8]. Forming dense objects by consolidating ceramic powders without pressure is a challenging task due to the unavoidable presence of pores and cracks that impair the mechanical performance. For the fabrication of dense ceramics, formulations with high solid loadings are required, which consequently restrict the suitability of slurries for the manufacturing of fine structures due to their decreased flow ability, increased viscosity and inhomogeneity. When shaping ceramic slurries with light, a significant mismatch between the refractive index of the ceramic powder and that of the photocurable resin strongly reduces the curing depth and additionally causes a coarsening of the curing resolution due to the side scattering of incident light [9]. Consequently, it is far more challenging to 3D print 'dark' ceramic materials ($Si_3N_4$, SiC) relative to alumina, silica or zirconia.

Polymer-derived-ceramics (PDCs) define a class of ceramic materials that are synthesized by thermal treatment (usually pyrolysis) of ceramic precursors (so-called preceramic polymers) under an inert or reacting atmosphere [10]. By using suitable preceramic polymers, various ceramic compositions such as amorphous silicon carbide (SiC), silicon oxycarbide (SiOC), and silicon carbonitride (SiCN), can be obtained after heat treatment at 1000-1100 °C in an inert atmosphere (argon or nitrogen). Because there is no sintering step, PDC parts can be formed without pressure at lower temperatures relative to traditional ceramic powder shaping technologies. Preceramic polymers can be processed using existing technologies suitable for polymers in general. Due to their distinctive nanostructure of carbon-rich and free carbon domains, PDCs show exceptional stability against oxidation, crystallization, phase separation and creep even up to 1500 °C [10].

Currently, PDCs have been successfully utilized for the fabrication of ceramic fibers, ceramic matrix composites (CMCs) and microstructures, e.g. microelectromechanical systems (MEMS). Since some preceramic polymers contain silyl and vinyl groups, they can act as negative photoresists, exploiting a UV activated hydrosilylation reaction. For this reason, photolithography is frequently used to achieve fine two-dimensional patterning in PDC MEMS [11–16]. However, UV activated hydrosilylation requires adding large amounts of photoinitiators (PI) and long exposures to high energy UV light, which makes its utilization unrealistic in additive manufacturing processes. To overcome this limitation certain polysiloxanes can be modified by the attachment of curable functional groups to the silicone backbone in order to render them UV active and facilitate rapid photopolymerization. The additive manufacturing of macro-sized complex SiOC parts with excellent mechanical strength and stability against high temperature oxidation by stereolithography of preceramic polymers was first reported in [17,18]. In the first approach, two silicones terminated with vinyl and mercapto functional groups were blended together to trigger cross-linking [18,19]. In the second approach [17], to obtain a photocurable polymer suitable for stereolithography, a commercial polysiloxane (Silres ® MK Powder) was modified by attaching functional acrylate groups to the silicone backbone. Despite significant developments in the field, the full potential of PDC stereolithography has not yet been realized. The range of readily photocurable preceramic polymers remains quite confined and has thus far been limited to polysiloxanes, consequently only SiOC ceramics have been additively manufactured in macro scale [2,17,18,20–22]. A recent study examined 3D printing of SiC by blending acrylate based photopolymers with SiC precursors [23]. However, oxygen contamination from the photopolymers is unavoidable and is likely to largely degrade the product material's thermomechanical properties. Another disadvantage of this route is the extremely low ceramic yield [20,23], as the mixed photopolymers are completely eliminated during pyrolysis.

Classical radical photopolymerization pathways rely mainly on the addition reaction of carbon double bonds in acrylate or methacrylate, and they have been widely used in photolithography and stereolithography processes. Acrylate or methacrylate based photopolymerization routes suffer from fundamental limitations, including high levels of polymerization-induced volume shrinkage and consequent stress development, and oxygen inhibition of the cross-linking reaction. An alternative pathway involves thiol-ene free radical addition, where SH groups undergo addition reactions with carbon-carbon double bonds. This polymerization route is considered as a "click reaction" and therefore exhibits the following characteristics [24]: (a) a high reaction yield (b) fast reaction rate, (c) insensitivity to solvent parameters (d) negligible volumetric shrinkage, and (e) insensitivity towards oxygen and water. These overall properties render thiol-ene click chemistry a promising approach towards photopolymerization in general and additive manufacturing in particular. It has been shown that the preceramic polymer polycarbosilazane, can be photopolymerized with thiol-ene reaction in 2D geometries and converted to silicon carbonitride based material upon pyrolysis [13].

In this work, we introduce a versatile stereolithographic route, based on VAT photopolymerization, using fast and inexpensive thiol-ene click reactions to 3D print Si-containing thermosets. These are subsequently pyrolyzed to obtain high performance Si-containing ceramics. Our approach is applied for different classes of pr-ceramic polymers containing carbon-carbon double bonds, including polysiloxane, polycarbosilane and polycarbosilazane with side vinyl groups. Due to the rapidity and efficiency of thiol-ene click chemistry reactions used here, the AM process can be effectively performed with conventional light sources (such as projectors) on benchtop Digital Light Processing (DLP) printers. Due to their high optical transparency, which minimizes the scattering effects, resin mixtures are further applicable for more sophisticated techniques such as two-photon polymerization [25,26] and microstereolithography [27].

## 2. Materials and methods

### 2.1 Materials:

The preceramic polymers used in this work are a liquid methylvinylhydrogen polysiloxane (polyramic SPR-212) and allylhydrydopolycarbosilane (StarPCS™ SMP10), bought from Starfire Systems (USA), as precursors for SiOC and SiC ceramics respectively. For SiCN, a liquid methylvinylhydrogen polycarbosilazane (Durazane 1800) bought from Merck, Germany) was selected. 1,6-hexanedithiol, phenylbis(2,4,6-trimethylbenzoyl)phosphine oxide (BAPOs), Sudan Orange G, and hydroquinone were purchased from Sigma Aldrich , Germany. All chemicals were used without further purification.

Formulation of photopolymerizable preceramic resins for stereolithography was conducted as follows: The preceramic resin was prepared by mixing preceramic polymers with phenylbis (2,4,6-trimethylbenzoyl) phosphine oxide (BAPOs) as photoinitiator (PI), Sudan Orange G as photoabsorber and hydroquinone as free radical scavenger. In order to homogenize the resin, it was treated in an ultrasonic bath for 2 hours, followed by degassing under vacuum. After that, 1,6-hexanedithiol (2T), which provides the thiol groups to react with the alkene functionalities of preceramic polymers, was dissolved in the resin mixture via mild stirring for 1 hour. The prepared resin was subsequently stored in a brown glass bottle to avoid exposure to light.

Upon exposure to light with wavelengths below 450 nm, which are included in the emitted white light from the projector, the PI undergoes a cleavage process generating free radicals (see Fig. S1 in SI), which then initiate the thiol-ene click reaction as seen in Fig. 1. Sudan Orange G, which exhibits suitable optical absorbance in the relevant working spectrum of the photoinitiator, was utilized to modify the sensitivity of prepared resin and therefore limit the penetration depth of the emitted light. As a result, the z-axis resolution can be controlled. Detailed information of resin sensitivity and critical exposure time can be found in Fig. S2 of the provided supplementary information. Hydroquinone acts as free radical scavenger to avoid any unwanted photopolymerization induced by background light or scattering, and was found to significantly prolong the shelf life of the resin.

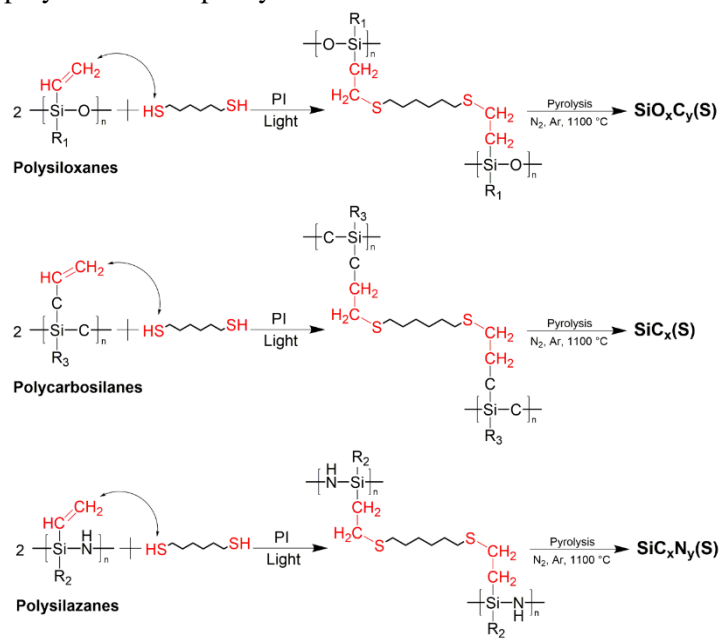

Fig. 1. Schematics of the photoinduced thiol-ene click reactions (cross-linking) of vinyl-containing Si-based preceramic polymers with 1,6-hexanedithiol via an anti-Markovnikov addition.

## 2.2 Stereolithography and thermal pyrolysis:

Stereolithography of preceramic polymer resins was conducted in an Open 3D Resin Printer LittleRP2 (LittleRP, USA) using an Acer X152H projector as a light source. The lens system of the projector was modified for short distance focusing. The z-axis resolution, namely the layer thickness of each print, was set at 100 µm to enable rapid printing. As a result, objects with 2 cm height could be fabricated in approximately one hour. The printing parameters including the exposure time are reported in Table S1. After printing, the parts fabricated from polysiloxane were washed with isopropanol, while those fabricated from polycarbosilane and polycarbosilazane were washed with anhydrous cyclohexane. Subsequently parts were post cured under exposure to UV irradiation between 350 and 400 nm for 30 mins. To transform these parts into ceramics they were pyrolyzed in a tubular furnace at 1100 °C for 2 hours under flowing nitrogen. The heating/cooling rate was set to 40 °C /h. A schematic illustration of the complete printing process is shown in Fig. 2.

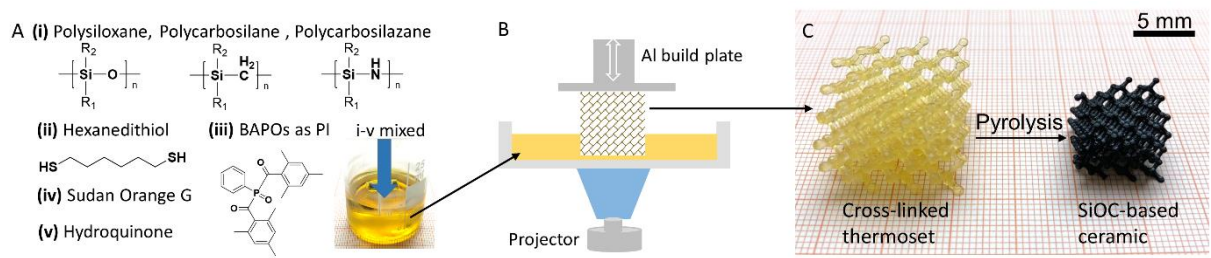

*Fig. 2. Additive manufacturing of ceramics from preceramic polymers: schematic representation of thiol-ene click chemistry based stereolithography. (A) The preceramic photopolymerizable resin was prepared by mixing different preceramic polymers (polysiloxanes, polycarbosilane, polycarbosilazane) with 1,6-hexanedithiol dithiol along with phenylbis (2,4,6-trimethylbenzoyl) phosphine oxide (BAPOs) as photoinitiator (PI), Sudan Orange G as photoabsorber and hydroquinone as free radical scavenger (for more details, see Materials and Methods). (B) Photopolymerization of preceramic resins in a benchtop DLP printer utilizing a commercially available projector as a light source. (C) A representative printed cross-linked thermoset and the resultant glassy ceramic formed upon pyrolysis at 1100 °C under $N_2$ atmosphere.*

## 2.3 Characterization:

Fourier transform infrared spectroscopy (FTIR) in Attenuated Total Reflection (ATR) mode carried out in an Equinox 55 (Bruker, Germany) in the range of 500 – 4000 cm$^{-1}$ was used to characterize the free radical initiated thiol-ene polymerization as well as the polymer to ceramic transformation. In-situ Differential Scanning Calorimetry (DSC) under exposure to UV light in the range between 350 and 600 nm (100 mW/s) was utilized in a DSC Q2000 (TA Instruments) to characterize the heat flow and reaction kinetics during photopolymerization. X-ray photoelectron spectroscopy (K-Alpha TM, Thermo Scientific) was used to determine the elemental composition of printed parts. The polymer to ceramic conversion was investigated in nitrogen with a heating rate of 10 °C min$^{-1}$ with thermal gravimetry/differential thermal analysis (TG/DTA) on STA 409 PC LUXX (Netzsch, Germany) coupled with a mass spectrometer OMNi Star GSD 320 (Pfeiffer Vacuum, Germany). Solid-state $^{29}$Si NMR spectra was recorded with a Avance 400 MHz spectrometer (Bruker, Germany) operating at 79.44 MHz to characterize the chemical environment of silicon in polysiloxane derived SiOC ceramics. Transmission electron microscopy (TEM) was used to prove the amorphous nature of SiOC. X-ray imaging (radiography) was performed using a microfocus X-ray source and a flat panel detector (Hamamatsu, Japan) with an area of ~ 120 x 120 mm² and a pixel size of 50 µm as reported in [28]. Sample magnification was set between 6 and 10, depending on the size of the sample, resulting in an acquired projection with a pixel size between 5 and 10 µm. For the tomographic acquisition 1000 projections were taken over a sample rotation of 360°. The printing quality and microstructure of 3D printed SiOC ceramics was evaluated by Scanning Electron Microscopy in Leo Gemini 1530 (Zeiss, Germany).

## 2.4 Mechanical properties:

Mechanical tests were conducted on SiOC honeycombs additively manufactured from polysiloxane SPR212 and pyrolyzed at 1100 °C for 2 hours under flowing nitrogen. Nanoindentation of SiOC honeycomb that had been embedded in epoxy resin and polished to a 1 µm finish, was conducted using a TI 950

TriboIndenter (Hysitron Inc., USA) with a tetrahedral Berkovich diamond tip at a load of 10 mN and a load rate of 50 µN/s. The elastic modulus and hardness are calculated by evaluating the slope of the curve dP/dh upon the elastic unloading with 200 indents. The compressive strength of SiOC honeycombs was determined using a universal test machine RetroLine (Zwick/Roell, Germany) with a cross-head speed of 0.1 mm/min. For this test, samples were first ground to achieve relatively parallel surfaces and a thickness of 2.78 mm. These were then sandwiched between two steel face sheets with epoxy glue to prevent horizontal movement. The solid cell wall density ($\rho_s$) of SiOC honeycombs was determined via Archimedes' method using distilled water, while the cellular density of this honeycomb ($\rho^*$) was calculated according to [29] by dividing the foam mass by its volume, which is the product of its length (a), width (b) and height (c).

## 3. Results and discussion

### 3.1 Thiol-ene click chemistry assisted stereolithography

To demonstrate the versatility of our approach we applied three commercially available Si-containing polymers from three different classes as precursors for three different ceramic compositions, i.e. polysiloxane SPR212 as precursor for SiOC, polycarbosilane SMP10 as precursor for SiC and polysilazane Durazane1800 as precursor for SiCN. The thiol-ene click reaction assisted AM process is schematically illustrated in Fig. 2. Preceramic polymers mixed with 1,6-hexanedithiol and BAPOs (PI) are cross-linked in less than 7 seconds using a commercial digital light processing (DLP) based projector.

Upon exposure to the white light, the PI undergoes a cleavage process generating free radicals that initiate the thiol-ene polymerization reactions shown in Fig. 1 in which thiol groups react with the alkene groups of preceramic polymers leading to a thioether linkage. Because of this propagating cross-linking, the liquid preceramic polymers transform into robust infusible thermosets that are essential for the preservation of 3D printed geometries during subsequent processing including washing and pyrolysis. Insufficient cross-linking would lead to soft structures that would deform prior to or during the ceramization process.

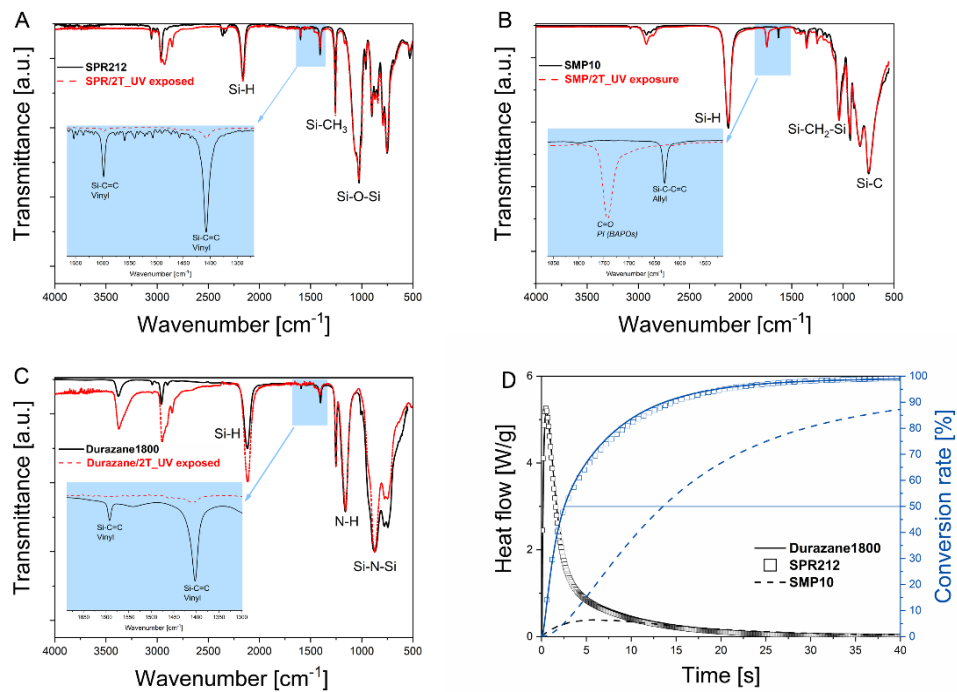

*Fig. 3. (A-C) ATR-FTIR spectra of preceramic resins before (black) and after photoinduced curing (red): (A) with polysiloxane SPR212, (B) polycarbosilane SMP10 and (C) polycarbosilazane Durazane1800. The thiol-ene click polymerization is confirmed by the strong reduction of the absorption bands of the vinyl-groups. (D) In-situ differential scanning calorimetric characterization of preceramic resins exposed to light demonstrating that the preceramic resins are effectively cured within 15 secs, which makes them suitable for stereolithographic additive manufacturing: (black) the curing reaction heat and (blue) the conversion rate (the integral curve of heat flow after normalization)*

.

The evolution of the thiol-ene click reaction can be observed using FTIR spectroscopy by monitoring the intensity of bands associated with vinyl groups at 1407 cm$^{-1}$ (=CH$_2$ scissoring) and 1597 cm$^{-1}$ (C=C stretch) and with thiol groups at 2570 cm$^{-1}$. Fig. 3 A-C display the FTIR spectra of SPR212, SMP10 and Durazane1800 polymers before and after the thiol-ene assisted photopolymerization. Because of the low concentration of thiols and relatively weak absorption associated with the S-H groups, the thiol peaks are not observed in the FTIR spectra even before the UV irradiation. Nonetheless, the significant decrease of the intensity of the vinyl absorption bands in the FTIR spectra as well as the elimination of the strong thiol odor indicate a successful thiol-ene click reaction in all polymers exposed to light. The appearance of the C=O absorption band in SMP10 based resin is induced from the addition of BAPOs. However, in-situ real-time calorimetric characterization under UV irradiation reveal significant differences between the polymers studied in this work.

As shown by the black lines in Fig. 3D, the polysiloxane SPR212 as well as polysilazane Durazane1800 polymers exhibit strong and sharp exothermic peaks immediately after UV exposure, which indicate very high curing efficiencies for these polymers. The normalized integral of these exothermic peaks shows that more than 50% of the resin mixture can be cross-linked within 5 secs, which is an ideal time frame for the AM process, where fast and efficient reaction kinetics are required. In contrast, the polymerization of the SMP10 polymer was significantly slower and was accompanied by smaller heat evolution. This is because SMP10 has far fewer vinyl groups compared to SPR212 and Durazane1800, which leads to a lower degree of photoinduced cross-linking. Furthermore, the SMP10 polymer absorbs light in the blue region of the visible spectrum, as evident by its intrinsic orange color, resulting in a slower photopolymerization process. This difference has implications for the ceramization process as discussed below. Nonetheless, all preceramic polymers studied here can be applied in the stereolithographic fabrication of complex thermoset parts as demonstrated in Fig. 4.

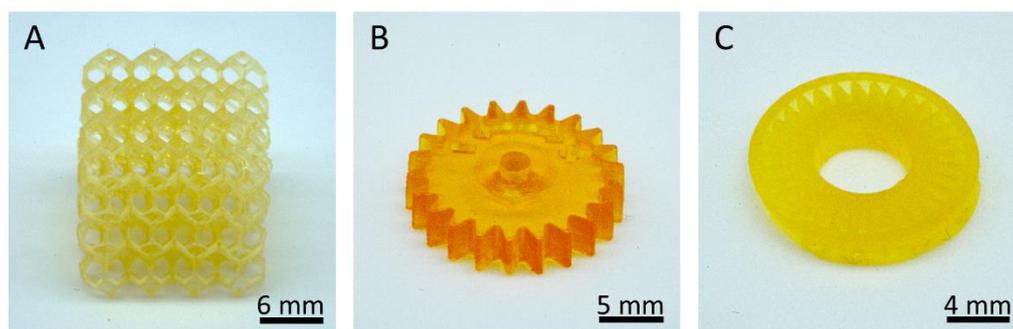

*Fig. 4. Representative examples of thermoset parts fabricated from three different polymer classes: (A) Kelvin cell structure printed with polysiloxane SPR212, (B) cog wheel printed with polycarbosilane SMP10 and (C) turbine structure printed with polycarbosilazane Durazane1800.*

### 3.2 Polymer-to-ceramic transformation and the microstructure of additively manufactured ceramic components

To transform the cross-linked thermosets into ceramics, they were subjected to further processing steps. First, they were rinsed with organic solvents (isopropanol for polysiloxane and anhydrous cyclohexane for polycarbosilane and polycarbosilazane), to remove the remaining non-polymerized resin, and post cured under UVA light between 350 – 400 nm to enhance their mechanical stability. In the next step they were transferred to a tube furnace and pyrolyzed in N$_2$ at 1100 °C. During pyrolysis the cross-linked preceramic polymers decompose into amorphous Si-containing ceramics; this transformation is accompanied by the release of hydrogen, methane and other carbon- and hydrogen-containing species. This process entails mass loss, isotropic volume shrinkage and densification of the material. Upon the pyrolytic conversion of cross-linked polymers to the corresponding ceramics, a linear shrinkage of roughly 30 % is observed (see Fig. 2C). Since the shrinkage is isotropic, the thermoset is transformed into a ceramic without any noticeable distortion. The preservation of geometry demonstrates that thiol-ene click reactions are an effective and easily accessible pathway towards the AM fabrication of ceramic parts with excellent resolution and smooth surfaces from preceramic polymers. To avoid cracking or explosion during pyrolysis, the geometry of printed preceramic polymers is designed with a feature size smaller than 2 mm in at least one direction. This reduces the diffusion path of evolving gas and facilitates the shrinking process.

Using the procedures developed in the present work, further ceramic compositions beyond the SiOC, SiC and SiCN materials studied here can be obtained by means of suitable preceramic polymers. General guidelines for the selection of appropriate ceramic precursors for use with thiol-ene click reactions can be stated as follows: (i) an inorganic-organic hybrid polymer or monomer, where the inorganic core comprises species (such as Al, B, Si, Ti or Zr) that form stable oxides, carbides, nitrides or oxycarbides upon pyrolysis; (ii) the presence of unsaturated carbon double or triple bonds, such as those in vinyl groups; (iii) the propensity to form thiyl radicals through photoinitiation. Furthermore, the methodology developed here can be easily adapted to produce composite ceramics containing fillers to impart catalytic, photocatalytic or magnetic functionalities.

In this work we were able to fabricate complete black glassy ceramics composed of silicon oxycarbide and silicon carbonitride. Fig. 5 A-D displays several examples of ceramic parts - ranging from cellular lattices to bulk structures - fabricated by our approach. Substantial internal flaws were not observed in corresponding radiography images (Fig. 5 E - H), which validates our approach towards AM of high performance ceramics. The SiOC and SiCN(O) parts were successfully fabricated with excellent resolution and very smooth, pore free surfaces. Surprisingly, SiC parts fabricated from the polycarbosilane SMP10 cracked and exploded during pyrolysis when the heating temperature reached 600 °C. At temperatures between 550 and 800 °C, SMP 10 transforms into an inorganic SiC based material accompanied by release of hydrogen ($H_2$) and methane ($CH_4$), as well as volumetric shrinkage, which indicates the polymer to ceramic conversion [30–32]. As discussed before, relative to SPR212 and Durazane 1800, the SMP10 polymer was cross-linked at a lower degree and more importantly its polymer structure contains greater amounts of silyl (Si-H) and carbosilane (Si-$CH_2$-Si) groups, leading to greater hydrogen gas evolution during pyrolysis. Therefore, the significant hydrogen and methane gas egression during the ceramization process of SMP10, along with shrinkage stresses, resulted in the formation and rapid propagation of cracks manifesting in the explosion of samples.

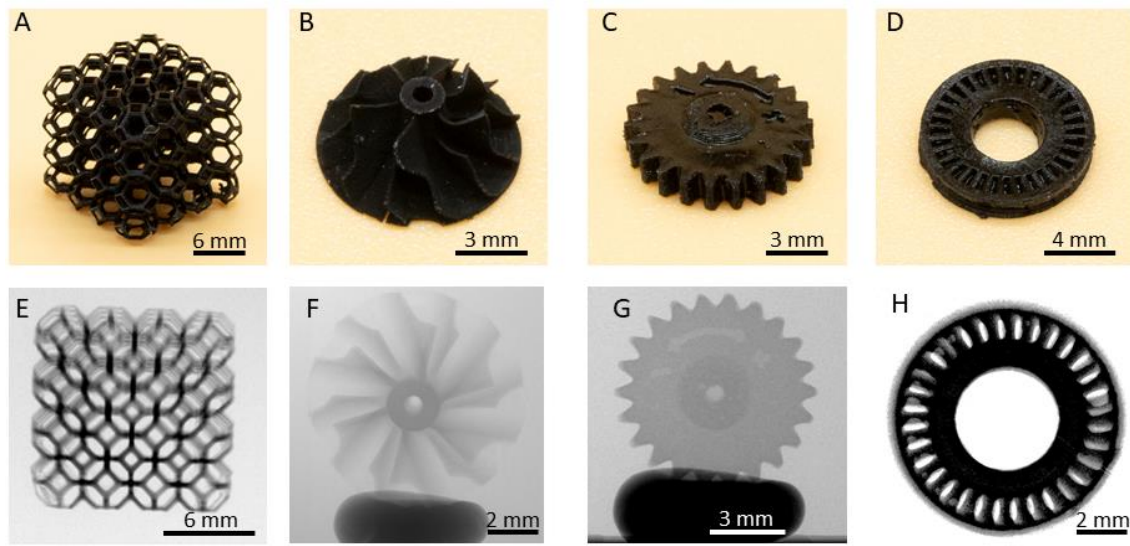

*Fig. 5. Examples of ceramic components fabricated by pyrolyzing the cross-linked thermosets showing (A,B) Kelvin cell structure and turbine impeller of SiOC and (C,D) cog wheel and turbine structure of SiCN(O). Corresponding radiography images (E-H) demonstrate the fabricated ceramic components are free from macro sized voids and defects.*

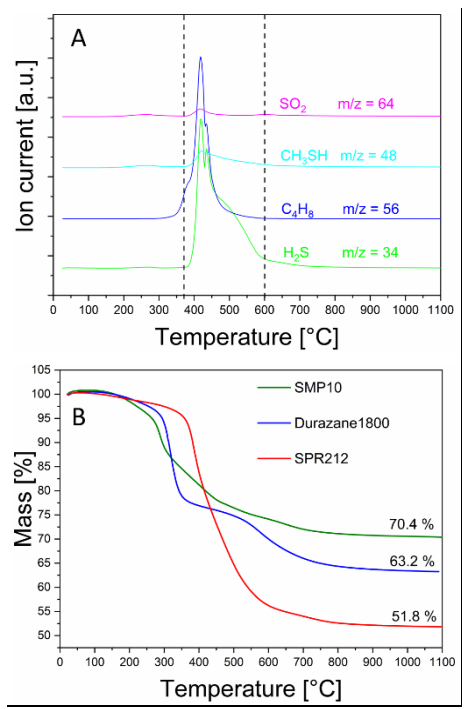

*Fig. 6. STA characterization of the polymer-to-ceramic transformation of the cross-linked thermosets fabricated from SPR212, SMP10 and Durazane 1800: (A) selected MS data for SPR212 indicating the release of $SO_2$ (m/z= 64), $CH_3SH$ (m/z= 48), $C_4H_8$ (m/z= 56) and $H_2S$ (m/z= 34) due to the decomposition of thioether linkages during the major mass loss region and (B) TG curves indicating the overall weight loss.*

The polymer to ceramic transformation was examined by thermal gravimetry coupled with mass-spectrometry, with results shown in Fig. 6. The overall weight loss of dithiol modified SPR212 resins is significantly larger in comparison to the weight loss of the unmodified preceramic polymer, while the differences in SMP10 and Durazane 1800 resins are less significant [33,34]. The additional weight loss is attributed to the evaporation and breakdown of thioether linkages during pyrolysis. As seen in Fig. 6 A, a substantial release of sulfur - containing low molecular weight species is observed between 400 and 600 °C. In contrast, in SMP10 and Durazane 1800 resins, oxidation in air leads to the formation of Si-O and Si-OH groups, which transform into thermally stable Si-O-Si bonds. The resultant ceramic yield is thus not strongly affected by the presence of dithiol compound.

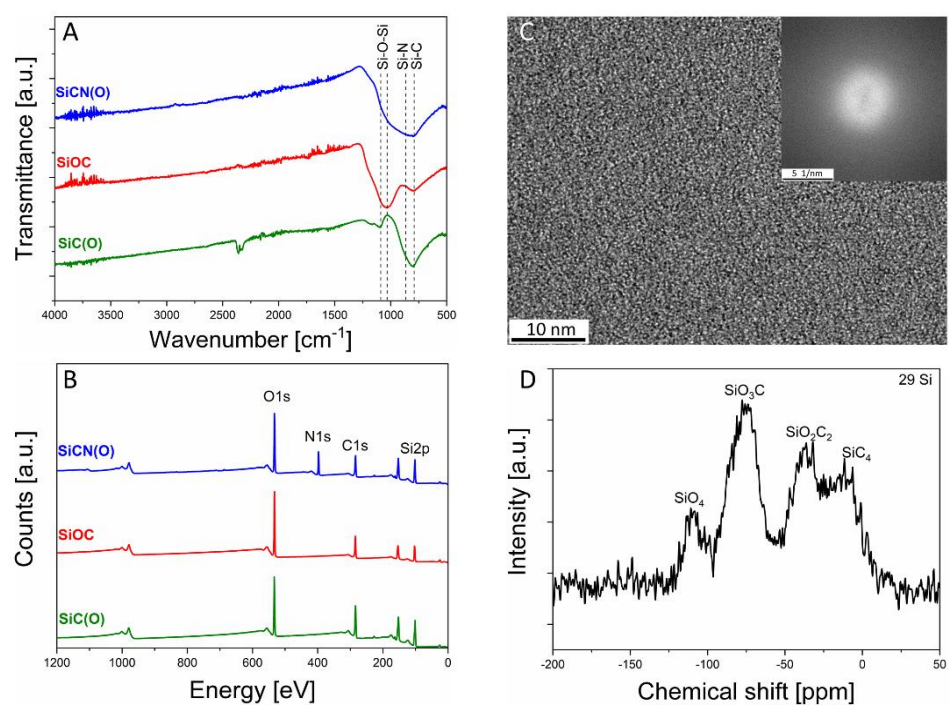

*Fig. 7. Characterization of additively manufactured polymer-derived ceramic materials: (A) ATR-FTIR and (B) X-ray photoelectron spectra of SiOC (red), SiC(O) (green) and SiCN(O) (blue line) synthesized at 1100°C from SPR212, SMP10 and Durazane 1800, respectively. Analysis of SPR212-derived SiOC ceramic pyrolyzed at 1100 °C: (C) TEM image, with selected area electron diffraction (SAED) pattern shown in inset indicating a homogeneous amorphous structure (D) solid state $^{29}Si$ NMR spectra, revealing mixed bonding between Si, O and C.*

*Table 1. Elemental composition of ceramics pyrolyzed at 1100 °C in flowing nitrogen*

| Preceramic polymer | Fabricated ceramic | Elemental composition, at. % | | | | |
|---|---|---|---|---|---|---|
| | | Si | O | C | N | S |
| Polysiloxane SPR212 | $SiC_{1.39}O_{1.53}S_{0.01}$ | 25.48 | 38.88 | 35.47 | 0 | 0.17 |
| Polycarbosilane SMP10 | $SiC_{1.22}(O_{0.78})S_{0.04}$ | 33.00 | 25.77 | 40.04 | 0 | 1.19 |
| Polycarbosilazane Durazane 1800 | $SiC_{1.02}N_{0.59}(O_{1.12})S_{0.04}$ | 28.52 | 28.92 | 26.22 | 15.27 | 1.07 |

As verified by FTIR and X-ray photoelectron spectroscopy and TEM characterization (Fig. 7), parts pyrolyzed at 1100 °C in flowing nitrogen are amorphous silicon-containing ceramics with a general formula of $SiC_xO_yN_z$ with small amount of residual sulfur. Their chemical composition depends on the preceramic polymer applied for the manufacturing (Table 1).

From XPS data, it is clear that ceramics with targeted compositions of SiC and SiCN possess significant oxygen content and these materials further demonstrate broad absorption bands around 1100 cm$^{-1}$ in the FTIR spectra attributed to Si-O-Si stretching vibrations. Since the additive manufacturing process was performed in ambient air, the silane (-Si-H) and silazane (Si-N-Si) bonds contained in the preceramic polymers reacted with atmospheric oxygen and moisture rapidly leading to an increased oxygen content. Consequently, "reacted" SMP10 and Durazane 1800 polymers are thermally converted into SiC(O) and SiCN(O) ceramics with deviated properties and microstructures compared to those in the intrinsic oxygen free SiC and SiCN ceramics. To minimize oxidation in stoichiometric SiC and SiCN ceramics, the entire fabrication process would have to be conducted under an inert atmosphere.

We thus focus on the representative SiOC ceramics fabricated from polysiloxane SPR212. As described above and as verified by TEM, the SiOC formed by pyrolysis at 1100 °C in flowing nitrogen is amorphous. Further insight into the SPR212-derived SiOC microstructure is provided by solid-state $^{29}$Si-NMR characterization (Fig. 7 D). The characteristic peaks at chemical shifts of -114 and -14 ppm correspond to $SiO_4$ and $SiC_4$ units and the peaks at -82 and -39 ppm can be assigned to mixed-bond $SiO_3C$ and $SiO_2C_2$ units [35]. The fabricated SiOC parts exhibited a measured Archimedes density of 2.10 g/cm$^3$, which approaches 97 % of the calculated theoretical density of. 2.17 g/cm$^3$ assuming densities of 2.2, 3.1 and 1.45 g/cm$^3$ for amorphous $SiO_2$, amorphous SiC and pyrolytic carbon, respectively [36]. This confirms that the fabricated SiOC parts are nearly fully dense with negligible pore volume. This is essential for ceramic materials, since cracks and pores act as stress concentrators under load and initiate failures. As can be seen in the radiography images (Fig. 5), SiOC components are free from observable internal cracks and closed pores. 3-D printed cellular cubes of SiOC are completely dense and free of macro sized defects as confirmed by the homogeneous absorption of X-rays in the tomographic images of the diamond lattice (Fig. 8 A) and Kelvin cell (Fig. 8 E) structures. The corresponding SEM images (Fig. 8 B-D and F-H) demonstrate again the delicate microstructures and fine features which are achievable only in additive manufacturing. Along the printing direction, step-like surfaces are clearly visible which is related to the layer-by-layer printing process. In this work the layer thickness during printing was set to be 100 µm to yield rapid fabrication. At higher magnification, glassy smooth surfaces are observed, exhibiting no porosity. This is consistent with the radiography images, indicating the high quality fabrication of ceramic components.

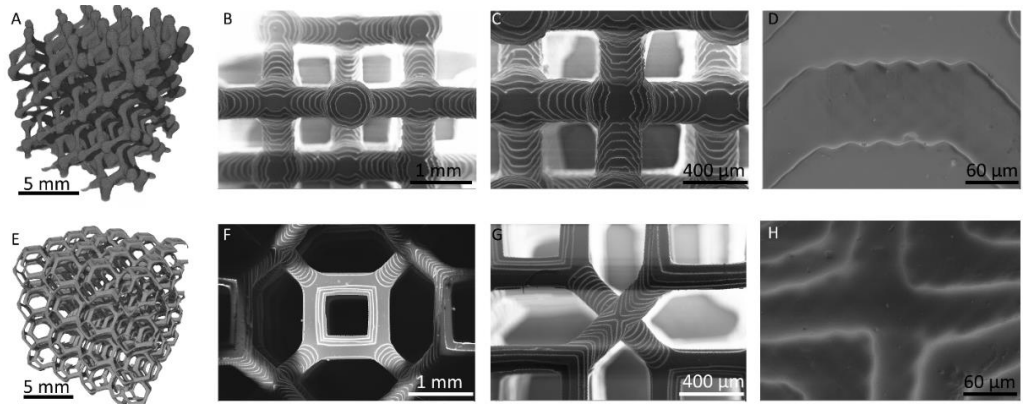

*Fig. 8. Morphology and microstructure of additively manufactured SPR212-dervied SiOC: (A and E) tomographic images of the diamond lattice (A) and Kelvin cell (E) structures. The delicate design is perfectly replicated without any detectable flaws. (B-D, F-H) Corresponding SEM images at higher magnification showing the step-like surfaces as well as the crack free struts.*

### 3.3 Mechanical properties

The hardness and elastic modulus of the SiOC ceramics were evaluated in nanoindentation experiments (Fig. 9). Fig. 9A shows the 3D printed honeycomb used for measurements along with an AFM image of an indent. A typical load-hold-unload curve is shown in Fig. 9B. 200 measurements were conducted to evaluate the reduced elastic modulus and hardness, as given in Fig. 9 C and D. Over a range of penetration depths, the mean values of the reduced Young's modulus and hardness are found to be 106±6 and 12±1 GPa, respectively. An earlier study showed that SiOC ceramics with lower O/C ratios behave similarly to amorphous-SiC materials, while the mechanical properties of SiOC with higher O/C ratio are similar to those in amorphous $SiO_2$[37]. In the present work, the hardness and elastic modulus of AM fabricated SiOC materials lie between the predicted limit of amorphous $SiO_2$ [37] and amorphous SiC [38] and are further in good agreement with reported values for silicon oxycarbide glasses with variable C/O ratio [37,39].

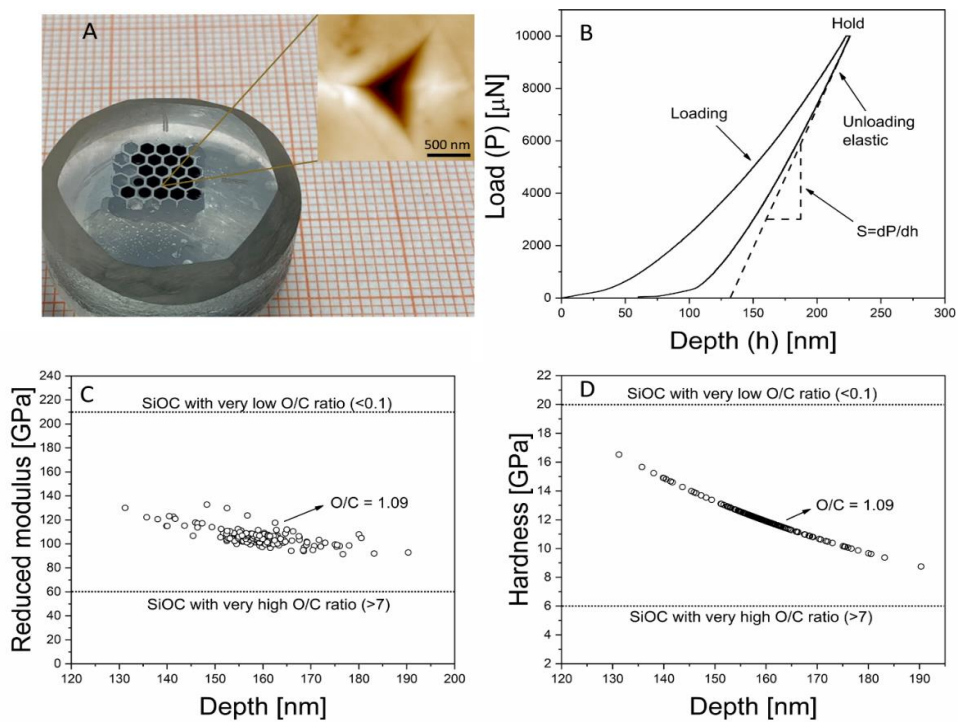

*Fig. 9. Results of nanoindentation tests performed on 2D honeycombs of additively manufactured SPR212-derived SiOC. (A) A polished SiOC honeycomb imbedded in epoxy resin. Inset shows the AFM image of one indent. (B) Load-Hold-Unload curve of one indent. Elastic modulus is determined via the elastic unloading process. (C, D) Reduced modulus and hardness as a function of the penetration depth. 200 measurement results are shown here.*

It is worth emphasizing that our fabrication method resulted in SiOC ceramics with comparable or even slightly better mechanical properties than those of SiOC glasses fabricated either via hot-pressing or via radio frequency (RF)-magnetron sputtering. Both of these methods yield high surface quality SiOC components (bar and thin films) with minimized defects and pore volume.

Since our manufacturing methods allow for the fabrication of 2D and 3D periodic cellular structures, in the following step we evaluate the out-of-plane compression strength of 2D SiOC honeycombs as a representative example of a material with high strength to density characteristics, and apply Gibson and Ashby's model [29] to analyze these results. The honeycomb geometry is particularly strong when loaded along the main axis of the hexagonal prism (out-of-plane, $X_3$) compared to the in-plane directions ($X_1$ and $X_2$), as thin cell walls are much stiffer under extension or compression than in bending. Therefore, honeycombs are mostly used as cores in sandwich panels to provide high strength to weight ratios. Likewise, honeycombs made from ceramic materials are frequently used as catalyst supports, filtration membranes and components in heat exchangers due to their superior chemical and mechanical stability in harsh environments.

Fig. S3 shows the corresponding stress-strain curves. 2D SiOC honeycomb structures fabricated in this work were found to exhibit a foam density of 0.61 g/cm$^3$ and a compressive strength of 216 MPa as seen in Table S2 and Fig. S3. The SiOC honeycomb demonstrates linear-elastic behavior as well as catastrophic failure, which is accompanied by a sudden drop in the measured stress. The plateau presented in the stress-strain curves before failure is resulted from slight misalignments of load bearing faces in the fabricated lattices, which in an ideal structure should be perfectly parallel, as well as an edge effect that, cell walls from internal to external can't fail simultaneously. Therefore, in our experiments the compressive strength is taken as the value at which this plateau occurs rather than the maximum crushing stress.

Gibson and Ashby's model relates the mechanical properties of a cellular solid to its relative density $\rho_{relative}$ (the density $\rho^*$ of the foam divided by the density $\rho_s$ of the solid which the foam is made of) and its solid strength/bulk modulus ($\sigma_s$, $E_s$). The strength ($\sigma$) as well as the modulus (E) of the cellular material can then be expressed as follows:

$$\frac{\sigma}{\sigma_s} = C \left(\frac{\rho^*}{\rho_s}\right)^m \quad (1)$$

$$\frac{E}{E_s} = C \left(\frac{\rho^*}{\rho_s}\right)^m \quad (2)$$

where $C$ is a dimensional constant and the exponent $m$ depends on the cell morphology. In the case of a 2D honeycomb carrying loads on the faces normal to $X_3$, equation (1) and (2) can further be simplified to (3) [40]:

$$\frac{E}{E_s} = \frac{\rho^*}{\rho_s} \quad and \quad \frac{\sigma}{\sigma_s} = \frac{\rho^*}{\rho_s} \quad (3)$$

In this case the Young's modulus of the 2D honeycomb simply reflects the solid modulus scaled by the load-bearing area. Using equation (3) with calculated relative density of 0.29 and the Young's modulus of 106±6 GPa from the nanoindentation experiments, we obtain 31±2 GPa for the theoretical honeycomb stiffness. Because of the measurement error, which occurred when evaluating the true strain of SiOC honeycomb under compression, we are unable to compare the measured Young's modulus with this calculated theoretical value here. More detailed explanation can be found in SI with Table S2 and Fig S4.

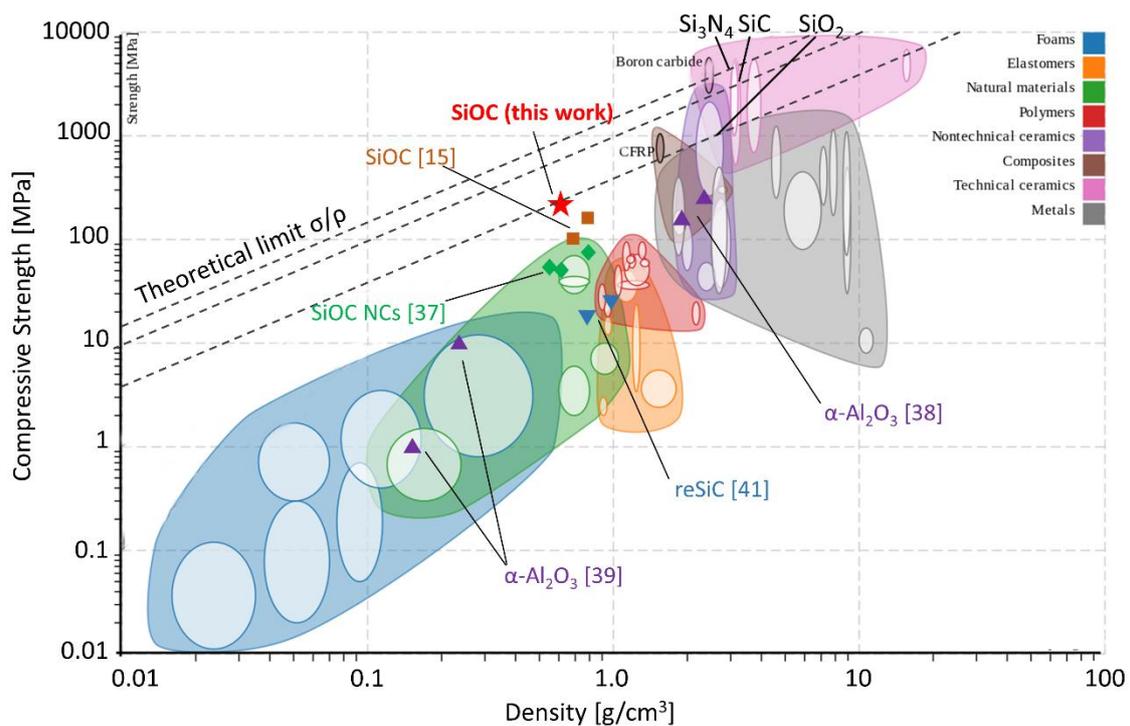

*Fig. 10. Strength-density Ashby chart for designing light strong structures. In this chart our additively manufactured SiOC honeycombs are compared to other ceramic (i.e. SiOC, Al$_2$O$_3$, SiC) honeycombs as well as to other natural and technical materials. Figure and material properties are generated by Nicoguaro using JavaScript and D3.js (https://commons.wikimedia.org/wiki/File:Material-comparison--strength-vs-density_plain.svg) except the data points of different ceramic honeycombs, which were inserted manually from the corresponding references.*

In Fig. 10, SiOC honeycombs printed in this work are compared, in terms of density and compressive strength, to natural and engineered cellular solids as well as other honeycombs made from SiOC [18], SiOC nanocomposites [41], Al$_2$O$_3$ [42–44], and SiC [45]. The silicon oxycarbide honeycombs fabricated in this work display an outstanding strength to density ratio that is significantly higher than other materials of similar density. Its absolute strength is also comparable with fiber-reinforced polymers and metal alloys. The compressive strength of 216 MPa is comparable to that exhibited by α-Al$_2$O$_3$ honeycombs of approximately twice of the density and is around one order of magnitude higher than commercial SiC honeycombs. The silicon oxycarbide honeycombs fabricated in this work further demonstrate a higher compressive strength compared to SiOC honeycombs with the densities ranged from 0.5 to 0.8 g/cm$^3$ reported in previous works (see e.g. [18] [41]). The exceptional performance of the additively manufactured SiOC honeycombs is the result of their dense and defect-free structure, which was observed by radiography and SEM characterization. As shown in equation (3), the strength of a cellular solid, whose cells walls display a stretch-dominated behavior, scales linearly with its relative density

[46]. This behavior is represented as a line with a slope of 1 in the double-logarithmic strength-density plot. Accordingly, the theoretical compressive stress of cellular polymer derived SiOC ceramics can be defined by these lines passing through bulk silicon carbide, silicon nitride and silica, which thus delineate the maximum obtainable performance for PDCs [47]. As can be seen, SiOC honeycombs in this work with a density of 0.6 g/cm$^3$ reach the theoretical compressive strength of silica based materials while approaching the predicted theoretical value of silicon carbide and silicon nitride. Shifting the ceramics' compositions towards higher C to O ratios by choosing polymers of higher carbon content and avoiding oxidation can further boost its specific strength towards the theoretical limit of SiC. By utilizing polysilazane in an oxygen free environment, silicon carbonitride ceramics can be obtained, offering further avenues towards high mechanical performance of printed cellular structures.

## 4. Conclusions

We have closely examined the utility of thiol-ene click chemistry towards additive manufacturing using preceramic polymers. We demonstrated a versatile stereolithographic manufacturing route whereby high performance polymer-derived ceramics are obtained with well controlled geometry. The developed approach can be applied for the entire range of known PDC systems. For precursors to be used in thiol-ene click assisted stereolithography they should exhibit an inorganic backbone, unsaturated carbon bonds and the photoinitiation of thiyl radicals. High levels of hydrogenated groups in precursors are detrimental as the evolution of a significant volume of hydrogen gas during pyrolysis can cause failure under certain firing conditions. The photocurable preceramic polymers of this method are easily prepared and can be applied in any SLA/DLP printer, as well as techniques like microstereolithography and two-photon polymerization to produce microstructures beyond the resolution limit of DLP techniques. The additively manufactured polysiloxane-derived SiOC components fabricated here are nearly fully dense, achieving 97 % of theoretical density. 2D SiOC honeycombs with a cellular density of 0.61 g/cm$^3$ exhibit a compressive strength of 216 MPa, surpassing the performance of comparable porous ceramics and stretching the boundaries of material property space in terms of strength to weight ratio under compression. The presently developed methods represent a flexible and rapid route towards high-performance additively manufactured polymer-derived ceramics that can find broad application in various engineering fields, particularly for load bearing materials applied in harsh environments and high temperatures. With appropriate AM techniques, ceramic components can be produced with the developed methods across a range of length-scales with high precision, representing a valuable new capability for industries such as aerospace, automotive, energy conversion and chemical engineering.

**Acknowledgements:**

The research was supported by the Deutsche Forschungsgemeinschaft under the grant "GU 992/17-1". We thank Jörg Bauer for helping with UV-DSC, Anke Maerten for conducting the nanoindentation measurements and Jan Epping for conducting the solid-state NMR experiments, respectively. We also thank Gaofeng Shao and Wuqi Guo for technical support in obtaining optical images.